\documentclass[
twocolumn,
aps,
prr,
citeautoscript,
showpacs,
showkeys,
groupedaddress,
superscriptaddress,
amsmath,
amssymb,
]{revtex4-2}

\usepackage{graphicx}
\usepackage{dcolumn}
\usepackage{bm}

\usepackage[hidelinks]{hyperref}
\hypersetup{
	colorlinks,
	linkcolor={blue!50!black},
	citecolor={blue!50!black},
	urlcolor={blue!80!black}
}

\usepackage{siunitx}
\usepackage{xcolor}
\usepackage{braket}
\usepackage{enumitem}

\usepackage{mathtools}
\newcommand{\Li}{\textsuperscript{6}Li }
\newcommand{\Linospace}{\textsuperscript{6}Li}

\usepackage{hyperref}

\begin{document}
	
\title{Floquet-Engineering of Feshbach Resonances in Ultracold Gases}

\author{Alexander Guthmann}
\affiliation{Department of Physics and Research Center OPTIMAS, RPTU Kaiserslautern-Landau, 67663 Kaiserslautern, Germany}

\author{Felix Lang}
\affiliation{Department of Physics and Research Center OPTIMAS, RPTU Kaiserslautern-Landau, 67663 Kaiserslautern, Germany}

\author{Louisa Marie Kienesberger}
\affiliation{Department of Physics and Research Center OPTIMAS, RPTU Kaiserslautern-Landau, 67663 Kaiserslautern, Germany}

\author{Sian Barbosa}
\affiliation{Department of Physics and Research Center OPTIMAS, RPTU Kaiserslautern-Landau, 67663 Kaiserslautern, Germany}

\author{Artur Widera}
\email[]{widera@rptu.de}
\affiliation{Department of Physics and Research Center OPTIMAS, RPTU Kaiserslautern-Landau, 67663 Kaiserslautern, Germany}

\date{\today}

\begin{abstract}
Scattering resonances are fundamental in science, spanning energy scales from stellar nuclear fusion to ultracold collisions. 
In ultracold quantum gases, magnetic Feshbach resonances have transformed quantum many-body research by enabling precise interaction control between atoms.
 Here, we demonstrate unprecedented control to engineer new Feshbach resonances at tunable positions via Floquet driving in a $^{6}$Li atom gas, achieved through strong magnetic field modulation at MHz frequencies. 
 This periodic modulation creates new resonances whenever dressed molecular levels cross the atomic threshold. 
 By adding a second modulation at twice the base frequency, we tune the asymmetry of resonance loss profiles and suppress two-body losses from Floquet heating. 
 This technique enhances control over atomic interactions, expanding possibilities for quantum simulations of complex systems and studies of exotic quantum phases.
\end{abstract}

\maketitle

When a quantum mechanical particle scatters from a potential, for certain incident energies the scattering cross section, summarizing the effect of the scattering event, shows a divergence. 
Such scattering resonances have profound consequences and are a ubiquitous phenomenon in nature, occurring across a broad range of energy scales and systems, including the Hoyle state in nuclear fusion within stars \cite{hoyle, Livio1989}, the production of the Higgs Boson \cite{Chatrchyan2012,Aad12}, Kondo resonances in solid-state systems \cite{doi:10.1126/science.1101077}, enhanced reaction rates in molecular collisions \cite{doi:10.1126/science.abl7257, paliwal2021quantum, doi:10.1126/science.adf9888}, and low-energy quantum systems such as ultracold atomic gases.
\\
At low collision energies, the microscopic interactions are governed by the $s$-wave scattering length $a_s$ \cite{RevModPhys.82.1225, BRAATEN20081770}.
For purely elastic scattering, $a_s$ is real, but in the presence of inelastic events, it becomes complex with an imaginary component \cite{hutson_feshbach_2007}.
The observation of magnetically tunable Feshbach resonances in a cold atom system \cite{inouye_observation_1998} sparked an explosion of research applications for ultracold quantum many-body systems, as it allows to change $a_s$ in both strength and sign through the external magnetic field.

\begin{figure*}
	\center
	\includegraphics{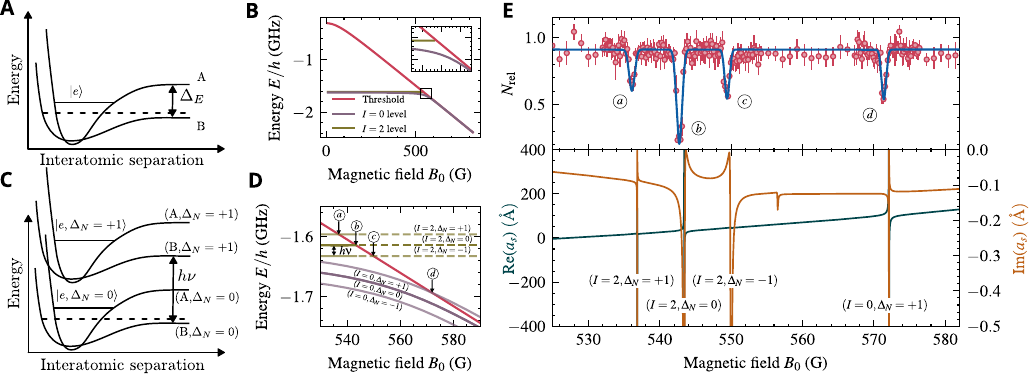} 
	
	\caption{\textbf{Occurance of Floquet-Feshbach resonances.} (A) Simplified sketch of the two-channel model for Feshbach resonances. An atom pair collides on potential B, coupled to potential A with a bound state $\ket{e}$. A Feshbach resonance occurs when the colliding pair is resonant with $\ket{e}$. If the potentials correspond to molecular configurations with different magnetic moments, their energy shift $\Delta_E$ can be tuned via the magnetic bias field. (B) Magnetic field dependence of the two highest molecular levels and the atomic threshold of two \Li atoms in the lowest hyperfine states $\ket{a}$ and $\ket{b}$. The two accessible $s$-wave Feshbach resonances correspond to molecular levels with total nuclear spin $I=2$ (narrow resonance at $\qty{543.28}{G}$) \cite{PhysRevLett.108.045304} and $I=0$ (broad resonance at $\qty{832.18}{G}$) \cite{PhysRevLett.110.135301}. At low collision energies, a Feshbach resonance occurs when a molecular state crosses the atomic threshold. Inset highlights the $I=2$ resonance. (C) and (D) Modulating the magnetic field dresses the molecular levels, and a Floquet-Feshbach scattering resonance occurs when a dressed molecular state crosses the atomic threshold. Arrows indicate resonance positions, with dashed lines indicating the continuation of the $I=2$ scattering pole into the continuum. (E) Floquet-Feshbach resonances at $\qty{18.15}{MHz}$ modulation with $B_\text{rf}^{(\text{a})} \approx \qty{3}{G}$. The upper panel shows experimental loss spectroscopy data, fitted with Gaussians to extract resonance positions. The lower panel displays the calculated real (blue) and imaginary (ochre) parts of the $s$-wave scattering length $a_s$, both showing resonant features corresponding to atom loss. Error bars represent standard deviation over at least five measurements per data point.}
	\label{fig:fig1}
\end{figure*}

This is accomplished by bringing a bound state of one molecular ground-state potential into resonance with another hyperfine molecular potential as schematically shown in Fig.~\ref{fig:fig1} (A).
One of the most significant applications of this tunability is in exploring the BEC-BCS crossover, where Feshbach resonances enable a smooth transition between a Bose-Einstein condensate (BEC) of tightly bound molecules and a Bardeen-Cooper-Schrieffer (BCS) superfluid of loosely bound Cooper pairs \cite{PhysRevLett.92.040403,PhysRevLett.92.120403,PhysRevLett.92.120401}.

The position of magnetic Feshbach resonances is determined by the magnetic field value where the energy of the bound state crosses the atomic threshold, as shown in Fig.~\ref{fig:fig1} (B) for \Linospace, and the width is dictated by the atomic properties of the species involved.
To date, magnetic Feshbach resonances have been observed for cold atomic \cite{inouye_observation_1998,RevModPhys.82.1225, Barbe2018, PhysRevA.102.033330, PhysRevX.5.041029, Thomas2018}, ionic \cite{Weckesser2021} and molecular \cite{Park2023, doi:10.1126/science.aau5322} systems.
However, the dependence on magnetic fields to tune the interaction strength imposes limitations, especially for atomic species with weak or inaccessible resonances. 

Furthermore, the relative position of Feshbach resonances in multi-component systems cannot be adjusted with a static magnetic field. 
To realize applications such as simulating color superfluidity \cite{10.1063/1.3001867} or \mbox{(Bose-)} Kondo physics in quantum gases \cite{Skou2021}, further control over resonance positions is crucial for adjusting intra- and inter-species interactions in multi-component systems.

\begin{figure*}
		\center
		\includegraphics{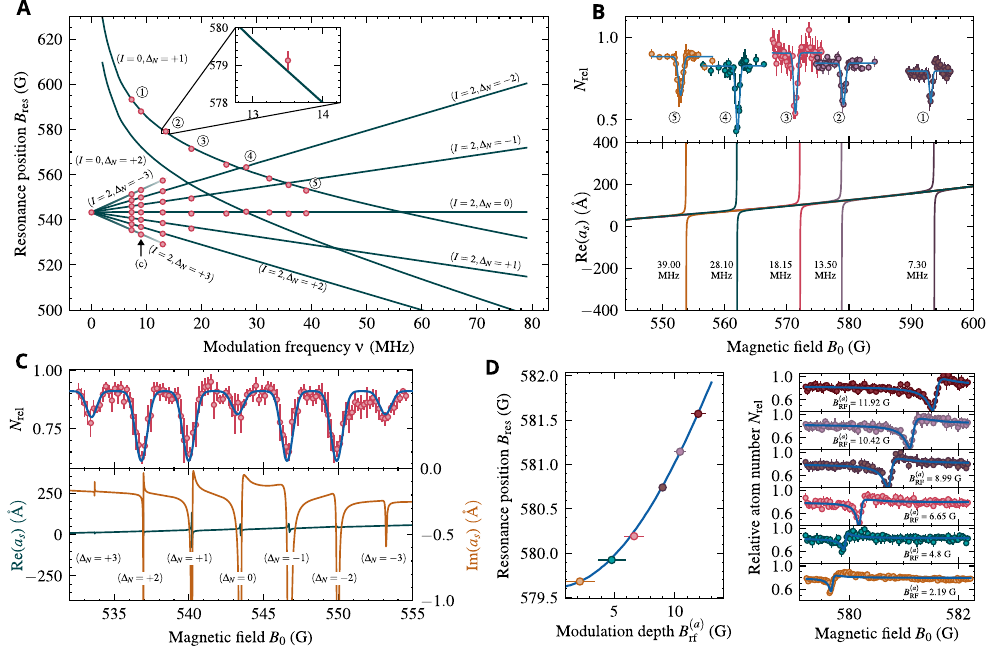}
		
		\caption{\textbf{Tuning of Floquet-Feshbach resonances.} (A) Floquet-Feshbach resonance positions as a function of modulation frequency. Solid lines show resonance position obtained from numerical coupled-channel calculations. Data points show measured resonance positions using loss spectroscopy. The error bars indicate $1 \sigma$ statistical error arising due to bias coil current fluctuations, with frequency uncertainty being negligible.
			(B) Magnetic field shift of the $(I=0, \Delta_N= + 1)$ Floquet-Feshbach resonance for five modulation frequencies. The upper panel shows experimental data fitted with Gaussians, while the lower panel shows theoretical scattering length calculations. This resonance shows the largest shift of $\qty{40.3}{G}$ between $\qty{7.3}{MHz}$ and $\qty{39.0}{MHz}$. (C) Loss spectrum and scattering length recorded at a modulation frequency of $\qty{9}{MHz}$ and a modulation strength of $B_\text{rf}^{(\text{a})} \approx \qty{5}{G}$ showing Floquet-Feshbach resonances of the $I=2$ level up to order $\pm3$. Error bars indicate $1 \sigma$ of observed variation in atom number $N_\text{rel}$. (D) Dependence of the $(I=0,\Delta_N=+1)$ resonance's position on the modulation strength for a driving frequency of $\qty{12.95}{MHz}$. The left panel shows calculated (solid line) and measured resonance positions. The right panel shows atomic loss spectroscopy data taken without artificial broadening of the resonance, exhibiting a strong asymmetry. The atomic loss is fitted with a Fano profile.}
		\label{fig:fig2}
	\end{figure*}

Advanced techniques extending the control of magnetic Feshbach resonances so far couple the open channel to other bound states by optical, microwave or radio-frequency (rf) transitions, where the AC stark-shift of the radiation field applied allows for limited shift of the resonance position.
Optical Feshbach resonances \cite{fatemi_observation_2000, PhysRevA.92.022709, PhysRevLett.115.155301}, where laser light is used to couple atomic states, allow for interaction control at the expense of rapid atom loss due to the short lifetimes of the excited state.
While such losses can be reduced using bound-bound transitions, it is restricted to suitable molecular transitions \cite{Bauer2009}.
Moreover, scattering parameters can be modified by coupling several bound states through rf radiation \cite{kaufman_radio-frequency_2009, hanna_creation_2010}, leading to shifts of existing resonances. 
\\
In this work, we implement a fundamentally novel approach that directly couples the scattering continuum to bound states, thereby inducing new resonances with an unprecedented level of control. This method has been proposed and theoretically studied using rf radiation polarized perpendicular \cite{tscherbul_rf-field-induced_2010, PhysRevA.81.041603, Beaufils2010, owens_creating_2016, PhysRevA.53.4343, PhysRevA.95.022709}, or by modulating the magnetic field parallel to the quantization axis \cite{smith_inducing_2015}, coupling the scattering continuum to bound states in different or the same spin channel, respectively. 
Recent theoretical work has further explored the origins and properties of such Floquet scattering resonances \cite{dauer2025understandingfloquetresonancesultracold}.
While offering precise control over resonance properties, experimental implementation has remained elusive due to the challenge of generating sufficiently large field amplitudes, which are so far only realized in the near-field regime on atomic chips to drive fast Rabi oscillations \cite{PhysRevA.110.053312}. 
In the case considered here, the rf modulation effectively dresses the molecular levels and potentials in a Floquet-type manner, as shown in Fig.~\ref{fig:fig1} (C) and (D), leading to the appearance of new resonances at magnetic field values whenever a dressed molecular level intersects the atomic threshold. 
Floquet engineering, successfully applied to tailoring band structures for cold atoms in optical lattices, has demonstrated powerful control over quantum states through periodic driving \cite{PhysRevLett.95.260404, PhysRevLett.99.220403, PhysRevX.9.041021, Weitenberg2021, Holthaus_2016}.
Here, we present experimental {realization and tuning of such Floquet-scattering resonances in a cold \Li gas through strong modulation of the magnetic field parallel to the quantization axis. 
	
By achieving modulation strengths comparable to the modulation frequency, we observe the formation of higher-order resonances. 

Notably, interference between different Floquet orders induces asymmetry in the loss profile of Floquet-Feshbach resonances.
Furthermore, we show that this asymmetry can be tuned by introducing a two-color driving scheme, where an additional driving term at twice the fundamental frequency allows us to suppress inelastic two-body losses associated with Floquet heating. 
By optimizing this two-color driving scheme, we achieve minimal losses, enabling the system to exhibit hydrodynamic behavior.
Conceptually, this two-color periodic modulation of the magnetic field corresponds to the first two terms of a Fourier series, with $B_\text{rf}^{(\text{a})}$ and $B_\text{rf}^{(\text{b})}$ defining the strengths of the first and second harmonics, respectively; further details are provided in the methods section.

\begin{figure*}
	\center
	\includegraphics{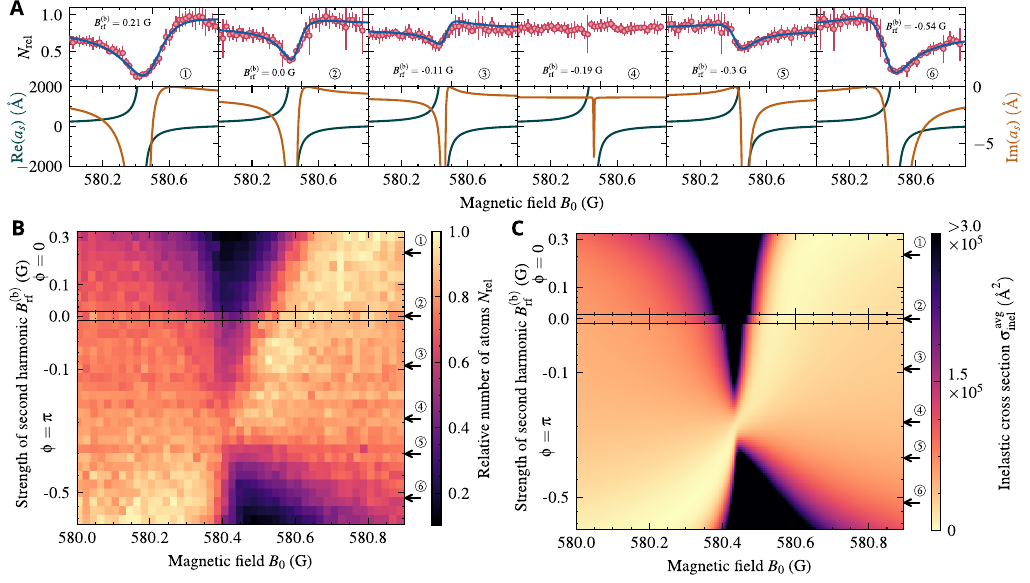}
	\caption{\textbf{Two-color engineering of Floquet-Feshbach resonances.} Effect of the additional two-color drive on the losses at the $(I=0,\Delta_N=+1)$ Floquet-Feshbach resonance. The strength of the first harmonic $B_\text{rf}^{(\text{a})}$ and the frequency $\nu$ are fixed at $\qty{8.9}{G}$ and $\qty{13.1}{MHz}$ respectively, while the strengths of the second harmonic $B_\text{rf}^{(\text{b})}$ is varied. (A) Atomic loss data and theoretical scattering length for six different strengths of $B_\text{rf}^{(\text{b})}$. The imaginary part of the scattering length is dramatically affected by the two-color drive, while the real part is affected only very slightly. The effect of this can be seen in the reduced atomic loss. (B) Loss data for 36 different values of $B_\text{rf}^{(\text{b})}$. The upper part shows the loss spectra for increasing $B_\text{rf}^{(\text{b})}$ with $\phi=0$. The lower part shows respectively the loss spectra for $\phi=\pi$, corresponding to negative $B_\text{rf}^{(\text{b})}$. (C) Thermally averaged inelastic scattering cross section $\sigma_\text{inel}^\text{avg}$ calculated using coupled-channel calculations, for the same range as in (B).}
	\label{fig:fig3}
\end{figure*}

\section{Tuning of Floquet-Feshbach resonances}

To experimentally study Floquet-Feshbach resonances we prepare a two-component \Li gas of an incoherent, equal mixture of the two lowest hyperfine states,  at a  temperature of $T< \qty{1.0}{\mu K}$.
On average, the number of atoms per spin state in the optical dipole trap is $2.3 \times 10^4$.
We perform atomic loss spectroscopy by applying the Floquet drive and vary the magnetic bias field $B_0$.
A resonance is detected by a drop in atom number, and its position is extracted through a fit. 
For the initial observation, an additional slow modulation ($600\,$Hz) of the magnetic field is added to artificially broaden the scattering resonances.
A typical loss spectrum obtained this way is shown in the upper panel of Fig.~\ref{fig:fig1} (E).

Several loss features can be seen which correspond to the position of resonance poles in the $s$-wave scattering length $a_s$, depicted in the lower panel.
In addition to resonance poles in the real part of the scattering length $a_s$ the imaginary part also shows an extremum at a resonance, leading to enhanced two-body losses at resonance. 
The microscopic process for such two-body losses is the absorption of a radiofrequency photon, which adds sufficient energy for the two colliding partners to leave the trap.
We label the resonances by the underlying dressed molecular state causing it, together with the relative number of drive quanta (Floquet order), e.g. $(I=0, \Delta_N=+1)$ labels the resonance caused by the $I=0$ level dressed by $\Delta_N=+1$ drive quanta.
\\
The magnetic position of Floquet-Feshbach resonances depends on the driving frequency $\nu$, as shown in Fig.~\ref{fig:fig2} (A). 
Two sets of resonances for \Li behave differently based on the underlying molecular state:
$I=2$ resonances depend linearly on frequency, as the dressed $I=2$ states are field-independent while the atomic threshold shifts linearly in the Paschen-Back regime.
This allows us to observe a ladder of Floquet resonances for small driving frequencies up to order $\Delta_N=\pm3$ as shown in Fig.~\ref{fig:fig2} (C). Interestingly, interference of different orders of the Floquet drive also modifies the static resonance, which is clearly attenuated in Fig.~\ref{fig:fig2} (C), and can be even extinguished.
In contrast, the Floquet resonances derived from the $I=0$ resonance exhibit a pronounced non-linear frequency dependence in Fig.~\ref{fig:fig2} (A), shifting the resonance by more than $250\,$G. 
This behavior is due to the strong magnetic field dependence of the $I=0$ molecular level, as shown in Fig.~\ref{fig:fig1} (B) and (D).
\\
The resonance position not only depends on the driving frequency but also on the driving strength.
Fig.~\ref{fig:fig2} (D) shows the positional shift of the $(I=0, \Delta_N=+ 1)$ resonance with respect to the modulation strength.
This shift is equivalent to an AC Zeeman effect and follows the relation $\Delta B_\text{res} \propto B_\text{rf}^2$ \cite{tscherbul_rf-field-induced_2010}. Unlike the AC Stark shift from an oscillating electric field, this is due to an oscillating magnetic field, with increasing modulation strength shifting the resonance to higher magnetic fields.
The loss data of Fig.~\ref{fig:fig2} (D) was taken without additionally broadening the resonance and therefore resolves the line shape.
A clear asymmetry in the loss profile can be seen well captured by a Fano profile \cite{PhysRev.137.A1364}, first observed in the context of photoionization spectra \cite{PhysRev.124.1866, PhysRev.137.A1364} and later found to be applicable to many phenomena \cite{PhysRevA.97.023411}.

\begin{figure*}
	\center
	\includegraphics{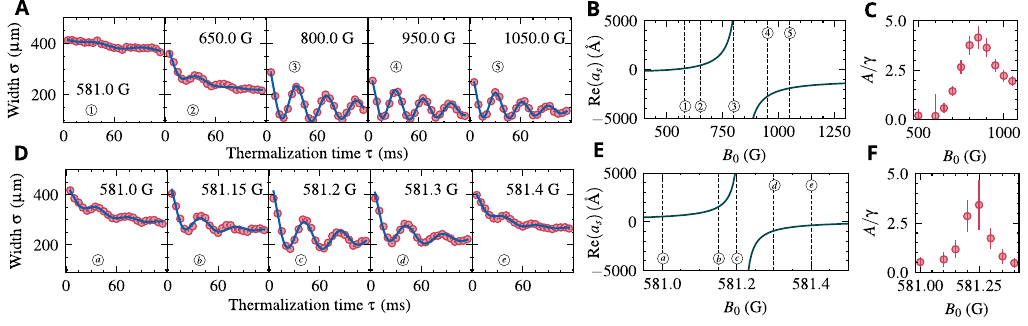}
	\caption{\textbf{Induced elastic interactions along a Floquet-Feshbach resonance.} (A) Evolution of the atomic cloud's width after a sudden quench of the trap depth for various magnetic fields across the static $I=0$ resonance, respectively in (D) across the $(I=0, \Delta_N=+1)$ Floquet-Feshbach resonance. The magnetic field values correspond to elastic scattering lengths as indicated in (B) and (E). By fitting the cloud's oscillatory response we can extract oscillation amplitude $A$ and decay rate $\gamma_A$. Plotting the ratio $A/\gamma_A$ over the magnetic field value shows typical resonance behavior, seen in (C) and (F). Errors bars give fit uncertainties. }
	\label{fig:fig4}
\end{figure*}

\section{Two-color drive}
While the inelastic imaginary part of the scattering length $a_s$ aids in detecting Floquet-Feshbach resonances, it also limits their application. 
These losses occur because two colliding atoms have an increased likelihood of absorbing drive quanta at resonance. 
However, the loss spectra in Fig.~\ref{fig:fig1} (E) show that the imaginary part is highly asymmetric, with distinct maxima and minima at specific magnetic fields. 
At the minima, the inelastic part of the scattering length becomes much smaller than the elastic part.
This asymmetry arises from interfering pathways in the Floquet-Hilbert space, where the magnetic field alters the phase between paths, resulting in constructive or destructive interference, resulting in the observed Fano profile.
\\
This interference suggests that adding more controllable pathways could further modify the asymmetry. 
We found that by introducing a second driving term at twice the fundamental frequency with variable strength $B_\text{rf}^{(\text{b})}$ and phase $\phi$, significant changes to the imaginary part of $a_s$ with minimal impact on the real part, as shown in Fig.~\ref{fig:fig3}, can be realized. 
Similar techniques have been studied for driven Hubbard systems \cite{PhysRevX.11.011057, PhysRevB.108.035151, chen2024, PhysRevResearch.4.013056}.
\\

For $\phi = 0$, increasing the amplitude $B_\text{rf}^{(\text{b})}$ shifts the loss minimum slightly towards higher magnetic fields on the attractive (BCS) side of the resonance, but increases losses on the repulsive (BEC) side. 
A more substantial effect occurs with a phase shift of $\phi=\pi$, corresponding to a sign inversion of $B_\text{rf}^{(\text{b})}$.
Here, the interference causes the loss minimum to shift across the entire resonance width, moving from the attractive to the repulsive side as $|B_\text{rf}^{(\text{b})}|$ increases.
At an intermediate value, the loss maximum nearly vanishes, leaving only a narrow symmetric feature at the resonance center.
This control over the asymmetry and position of the loss minimum allows for effective suppression of inelastic losses across both sides of the resonance, enhancing stability for applications requiring precise interaction dynamics.

\section{Elastic Interactions}
Building on this enhanced control over losses, we next confirmed the enhancement of elastic interactions at a Floquet-Feshbach resonance, which becomes more prominent as inelastic losses are suppressed. 
To explore this effect, we tracked cloud dynamics following a quench of the optical dipole trap depth. 
We employ the ensuing oscillatory dynamical response as a measure for elastic interactions.
Similar techniques have been used to explore collective oscillations in the BEC-BCS crossover \cite{PhysRevLett.92.203201, PhysRevLett.98.040401}
After preparing a cold gas, we simultaneously switch on the rf and quench the optical dipole trap to $20\%$ of its initial depth. 
The case of the static $I=0$ Feshbach resonance serves as a reference, and we extract the dynamics of the cloud width along the axial direction of the optical dipole trap to reveal the impact of interactions. 
This includes thermalization as the trap depth is reduced, but also the excitation of low-lying collective excitations driven by elastic interactions. 
Far from resonance, where interactions are weak, the cloud shows neither thermalization nor periodic dynamics.
However, close to resonance in both the BEC and BCS regimes, we observe pronounced oscillations in cloud width (Fig.~\ref{fig:fig4} (A)), indicative of collective hydrodynamic behavior resulting from elastic interactions.

To induce elastic collisions at $\qty{581}{G}$, where no such interactions are observed in the static case, we tune the modulation frequency to shift the Floquet-Feshbach resonance to this field. 
We induce elastic interactions at different magnetic fields around the Floquet-Feshbach resonance position, while simultaneously minimizing the inelastic losses by the two-color scheme for precisely this magnetic field value.
Thereby, we can induce similar oscillations at magnetic field values where the cloud previously showed no elastic response (Fig.~\ref{fig:fig4} D-F).
Comparing the oscillatory response between the static $I=0$ resonance and the Floquet-Feshbach resonance with two-color driving, we found similar dynamics, confirming that elastic scattering is enhanced under Floquet modulation, while the Floquet-induced loss and heating can be reduced to a level where still weakly-damped trap dynamics can be observed.

\section{Outlook}
In future studies it will be intersting to investigate resonance interference. When two resonances approach each other, they undergo an avoided crossing, potentially forming a bound state in the continuum \cite{friedrich_interfering_1985}.
Additionally, the rapid switching of the Floquet drive offers a promising tool for exploring quenched interactions and non-equilibrium dynamics in quantum gases. 
We also plan to study three-body losses and to create a degenerate quantum gas on both sides of a resonance, paving the way for deeper insights into many-body physics and phase transitions in Floquet-driven quantum systems.

\paragraph*{Acknowledgements}
This work was supported by the German Research Foundation (DFG) via the Collaborative Research Center Sonderforschungsbereich SFB/TR185 (Project 277625399), and we thank for many fruitful discussions with S. Eggert, A. Pelster, H. Ott and C. Dauer from which helpful ideas emerged.
We also thank A. Eckardt, I. Lesanovsky and J. Hecker Denschlag for carefully reading and commenting on the manuscript.

\paragraph*{Author contributions.}
A.W. and A.G. conceived the research. 
A.G. took and analyzed the experimental data and calculated the theoretical estimations. 
F.L., L.M.K. and S.B. helped run the experimental apparatus and collect data. 
A.W. supervised the research. 
All authors contributed to the interpretation of the data, writing of the manuscript, and critical feedback.

\section{Methods}

\subsection{Experimental Sequence}
We begin by preparing a cold gas of $^6$Li in a mixture of the two lowest hyperfine states, $\ket{a}=\ket{f=1/2, m_f = +1/2}$ and $\ket{b}=\ket{f=1/2, m_f = -1/2}$, in a steel vacuum chamber \cite{gaenger2018}. 
The gas is then transported via a dipole trap with tunable focal length into a glass cell to achieve the required modulation strength for studying Floquet-Feshbach resonances.
After transport, a perpendicularly polarized rf pulse is applied to equalize the mixture of states $\ket{a}$ and $\ket{b}$, followed by evaporative cooling at a bias magnetic field of $\qty{763}{G}$ to further lower the temperature and increase the density. 
This process results in approximately $2.3 \times 10^4$ atoms per spin state and a density of $n_0 = 4 \times 10^{14} \text{m}^{-3}$ inside the optical dipole trap.
At this stage, the axial and radial trap frequencies of the cigar-shaped optical dipole trap are $\omega_{z}=2 \pi \times \qty{68}{Hz}$ and $\omega_{r}=2 \pi \times \qty{14.3}{kHz}$, respectively.
The temperature is chosen to be low enough to be firmly in the $s$-wave dominated regime as well as to limit resonance broadening due to a thermal collision energy distribution.
We do not evaporate further to ultracold temperatures, since this would lead to molecule formation because of the underlying $I=0$ molecular state. 
We detect Floquet-Feshbach resonances using atomic loss spectroscopy.
The magnetic field is rapidly ramped to the target bias field $B_0$, held for $\qty{200}{ms}$ to stabilize, and then subjected to rf modulation. 
For the loss spectra shown in \autoref{fig:fig1} (E) as well as \autoref{fig:fig2} (A), (B) and (C) the bias field is additionally modulated at $\qty{600}{Hz}$ with $\approx \qty{0.7}{G}$ in amplitude, broadening the resonances for easier detection. 

The duration of the modulation is adjusted based on its strength to optimize the signal-to-noise ratio, \autoref{tab:modulation_strength_length} lists the times used to obtain the data of \autoref{fig:fig2} (D).

After modulation, the magnetic field is ramped back to $\qty{763}{G}$, and the number of remaining atoms is measured via absorption imaging.
We calibrate the magnetic bias field $B_0$ using rf spectroscopy, driving transitions between state $\ket{b}$  and $\ket{c}=\ket{f=3/2, m_f = -3/2}$. 
The loss spectroscopy measurements cycles are interleaved with cycles performing rf spectroscopy to account for magnetic field drifts. 

For the thermalization measurements the gas was prepared in the same way as described.
After the magnetic field has been allowed to settle at the target value, the trap was instantaneously ($<\qty{1}{us}$) quenched from $\qty{1}{W}$ to $\qty{200}{mW}$ and at the same time for measurements on the Floquet-Feshbach resonances, the two-color drive was switched on.
The amplitude of the second harmonic has been adjusted for minimal losses.
The gas is held for a certain time $\tau$ with or without the Floquet-drive applied, after which the gas is immediately imaged using absorption imaging at the current magnetic field.

\begin{table}[h]
\centering
\caption{Length of the applied Floquet modulation for the loss spectroscopic data in \autoref{fig:fig2} (D). }
\begin{tabular}{ |c|c| }
 \hline
 Modulation strength $B_\text{rf}^{(\text{a})}$ (G) & Modulation length (ms)  \\ 
 \hline
 $2.19$ & $1500$  \\ 
 $4.80$ & $300$  \\ 
 $6.65$ & $200$  \\ 
 $8.99$ & $100$  \\ 
 $10.42$ & $60$  \\ 
 $11.92$ & $40$  \\ 
 \hline
\end{tabular}
\label{tab:modulation_strength_length}
\end{table}

\subsection{Scattering Hamiltonian}
The total Hamiltonian $\bm{H}(t)$ of two scattered atoms in the center of mass frame can be expressed as
\begin{equation}
	\bm{H}(t) = \bm{T} + \bm{H}_{\text{hyp}} + \bm{H}_{\text{ee}} + \bm{H}_{\text{zee}}(t)\:,
	\label{eq:td_hamiltonian}
\end{equation}
with the hyperfine contribution represented by
\begin{equation}
	\bm{H}_{\text{hyp}} = \xi_A \vec{\bm{s}}^{(A)} \vec{\bm{i}}^{(A)} + \xi_B \vec{\bm{s}}^{(B)} \vec{\bm{i}}^{(B)} \:,
\end{equation}

where $\xi_i$ denotes the hyperfine constant for the $i$-th nucleus, and $\vec{\bm{s}}^{(i)}$ and $\vec{\bm{i}}^{(i)}$ represent the spin and nuclear spin of the $i$-th atom, respectively.
In the case of two \Li atoms $\xi_A = \xi_B = \xi$ and $\xi/h = \qty{152.1368407}{MHz}$ \cite{RevModPhys.49.31}.
$\bm{T}$ represents the usual kinetic energy operator in the basis of spherical harmonics.
The electronic interaction between the outer electrons of two alkali atoms is described by the term
\begin{equation}
	\bm{H}_{\text{ee}} = V^0(r) \bm{P}^0 + V^1(r) \bm{P}^1 \: ,
\end{equation}
where $V^0(r)$ and $V^1(r)$ are the singlet and triplet potentials, while $\bm{P}^0$ and $\bm{P}^1$ are the corresponding projection operators \cite{PhysRevB.38.4688, PhysRevA.78.052703}.
We consider the time-dependent magnetic field,
\begin{equation}
    B(t) = B_0 + B_\text{rf}^{(\text{a})}\cos{2 \pi \nu t} + B_\text{rf}^{(\text{b})}\cos{(4 \pi \nu t + \phi)},
    \label{eq:fourier_series}
\end{equation}
where $\nu$ is the modulation frequency, $B_\text{rf}^{(\text{a})}$ and $B_\text{rf}^{(\text{b})}$ are the first and second harmonic amplitudes, and $B_0$ is the static bias field. 
This time-dependent magnetic field causes the Zeeman operator
\begin{align}
	\label{eq:zee2}
	\bm{H}_{\text{zee}}(t) &= \mu_B \left(B_0 + B_\text{rf}^{(\text{a})}\cos{2 \pi \nu t} + B_\text{rf}^{(\text{b})}\cos{(4 \pi \nu t + \phi)} \right)  \nonumber\\
	& \quad \cdot \left( g_e \bm{s}_z^{(A)} + g_n \bm{i}_z^{(A)} + g_e \bm{s}_z^{(B)} + g_n \bm{i}_z^{(B)}  \right)  \: ,
\end{align}
which accounts for the interaction between the magnetic field and the magnetic moment of the system, to be time-dependent.
Where $\mu_B$ is the Bohr magneton, $g_e$ and $g_n$ are the electron and nuclear g-factors, and $\bm{s}_z^{(i)}$ and $\bm{i}_z^{(i)}$ represent the $z$-component of electron spin and nuclear spin for the $i$-th atom, respectively.
The dipole-dipole interaction for \Li is very weak and therefore neglected in our calculations.
Using Floquet theory this time-dependent periodic Hamiltonian can be transformed into an equivalent time independent Hamiltonian acting on a larger state space including periodic functions \cite{shirley_solution_1965, Holthaus_2016}.
In this representation the rf field is accounted for by
\begin{equation}
	\bm{H}_\text{rf} = \hbar \omega \left( \bm{N} - N_0 \right) \:,
	\label{eq:cum2}
\end{equation}
where $\bm{N}$ is the operator giving the number of drive quanta, while $N_0$ is an arbitrarily large positive offset.
Considering only $\phi=0$ and $\phi=\pi$, the atom-rf coupling can be written as 
\begin{align}
	\bm{H}_\text{cpl} &=  \frac{\mu_B}{2}  \Big\{ B_\text{rf}^\text{(a)} \left( \bm{\alpha} + \bm{\alpha}^\dagger \right) + B_\text{rf}^\text{(b)} \left( \bm{\alpha}^2 + (\bm{\alpha}^\dagger)^2 \right) \Big\} \nonumber \\  
    & \quad \cdot \left( g_e \bm{s}_z^{(A)} + g_n^{(A)} \bm{i}_z^{(A)} + g_e \bm{s}_z^{(B)} + g_n^{(B)} \bm{i}_z^{(B)}  \right) \: ,
\end{align}
where $\bm{\alpha}^\dagger$ and $\bm{\alpha}$ are creation and annihilation operators of drive quanta.
$B_\text{rf}^\text{(b)}$ being positive for $\phi=0$ and negative for $\phi=\pi$.
With this the total Floquet-Hamiltonian is given as
\begin{equation}
	\bm{\Tilde{H}} = \bm{T} + \bm{H}_{\text{hyp}} + \bm{H}_{\text{zee}}^\text{(0)} + \bm{H}_{\text{ee}} + \bm{H}_\text{rf} + \bm{H}_\text{cpl}\:,
\end{equation}
where $\bm{H}_{\text{zee}}^\text{(0)}$ captures the static part of \autoref{eq:zee2}.
This time-independent Floquet-Hamiltonian is then expressed in the uncoupled and anti-symmetrized basis \cite{PhysRevA.78.052703}
\begin{align}
	&\big\{ 2 ( 1 + \delta_{msA,msB} \delta_{miA,miB}) \big\}^{-\frac{1}{2}} \nonumber \\
     &\cdot \big\{ \ket{L,M_L;m_{sA},m_{iA};m_{sB},m_{iB}, N} \nonumber \\
     &- (-1)^L \ket{L,M_L;m_{sB},m_{iB};m_{sA},m_{iA}, N} \big\}\:,
     \label{eq:basis}
\end{align}
before being diagonalized using the for scattering problems well suited method of coupled-channel calculations.
The action of $\bm{\alpha}^\dagger$ and $\bm{\alpha}$ on this Basis is to raise and lower the Floquet order by one
\begin{equation}
    \bm{\alpha}^\dagger \ket{\rho, N} = \ket{\rho, N + 1}, \quad \bm{\alpha}\ket{\rho, N} = \ket{\rho, N - 1},
\end{equation}
where $\rho$ indexes all other quantum numbers of \autoref{eq:basis}.
Similarly the number operator $\bm{N}$ of \autoref{eq:cum2} is defined as
\begin{equation}
    \bm{N} \ket{\rho, N} = N \ket{\rho, N} \:.
\end{equation}
Floquet theory involves an infinite number of Floquet channels for a complete description of the system. 
However, in numerical calculations, the Floquet space must be truncated, and only a finite number of channels can be considered. 
If the Basis of \autoref{eq:basis} is truncated to Floquet orders from $N_0-\Delta_\text{max}$ to $N_0+\Delta_\text{max}$.
Then, a useful empirical guideline for selecting $\Delta_\text{max}$ is based on the ratio between the energies of modulation strength $\mu_B B_\text{rf}^\text{(a)}$ and the driving frequency $h \nu$, 
\begin{equation}
    \Delta_\text{max} \geq 3 \frac{\mu_B B_\text{rf}^\text{(a)}}{h \nu} \: .
\end{equation}
This criterion generally ensures sufficient convergence. However, it should be applied cautiously, and convergence must be verified manually for each specific case to ensure accuracy.

\subsection{Atom number measurement}
The number of atoms is determined using absorption imaging on the $\mathrm{D}_2$ line of \Li.
The cloud is imaged onto a sensitive sCMOS device with magnification $M$.
From those images the optical density
\begin{equation}
    \text{OD}(x,y) = - \log{\frac{I(x,y)}{I_0(x,y)}}
\end{equation}
is determined where $I$ is the intensity with, and $I_0$ the intensity without atoms present.
From this the number of atoms $N$ can be calculated by summing of all pixels
\begin{equation}
    N = \frac{A}{M \sigma_0} \sum_\text{pixel} \text{OD}(x,y)
\end{equation}
where $A$ is the pixel area size and $\sigma_0$ is the total absorption cross-section.
Fluctuations of the laser power, camera noise and varying loading and transport efficiencies  lead to fluctuating atom number measured for each experimental cycle.
Limited knowledge of the absorption cross-section leads to a systematic uncertainty in the total number of atoms present. 
Since we only need the relative number of atoms for our analysis, this issue is of no concern to us.
By comparing the number of atoms with Floquet drive applied $N_\text{w}$ to the number of atoms remaining without the Floquet drive enabled $N_\text{w/o}$  we get the relative number of atoms $N_\text{rel} = \frac{N_\text{w}}{N_\text{w/o}}$.

\subsection{Coupled-Channel Calculations}

The coupled-channel method solves the differential or equivalent integral equations for the radial coordinate, while the internal coordinates are diagonalized using a given basis set.
Because of improved stability and ease of including the appropriate boundary conditions we implemented a scattering code to solve the set of coupled Lippman-Schwinger integral equations using a spectral expansion into Chebyshev polynomials \cite{Rawitscher_2005, GONZALES1997134, GONZALES1999160}.  

A requirement for accurate coupled-channel scattering calculations is the knowledge of very good molecular potentials.
The potentials used for calculations of this work are based on published analytical potentials derived from spectroscopic data \cite{le_roy_accurate_good, DATTANI2011199}.
These potentials were further improved by iteratively adjusting the potential, with the correction term described in Ref. \cite{julienne_contrasting_good}, till the position of the $I=2$ Feshbach resonance matches the experimentally known value \cite{PhysRevLett.108.045304}.
For the numerical calculations of the AC Zeeman shift in \autoref{fig:fig2} and two-color drive in \autoref{fig:fig3} the potentials were additionally optimized to reproduce the measured resonance position for a modulation with frequency $\nu=\qty{13.1}{MHz}$ and strength $B_\text{rf}^{(\text{a})} = \qty{8.9}{G}$ at a collision energy of $\qty{724}{nK}$.
\\
The inelastic scattering cross-section of \autoref{fig:fig3} (c) is a thermal average over the Boltzmann distribution $p(E_c)$ of collision energies $E_c$ \cite{PhysRevX.10.011018}.
The thermally averaged inelastic cross section is then given as
\begin{equation}
    \sigma_\text{inel}^\text{avg} = \int p(E_c) \sigma_\text{inel}(E_c) \, \mathrm{d} E_c
\end{equation}
where the cross section for a specific energy is calculated from the scattering length by \cite{10.1063/1.2752162}
\begin{equation}
    \sigma_\text{inel}(E_c) = \frac{4 \pi \text{Im}(a_s)}{k(1 + k^2 |a_s|^2+ 2 k \text{Im}(a_s))} \: .
\end{equation}

\subsection{Fitting the loss spectroscopic data}
\label{sec:fitting_functions}
In case the additional slow modulation is used the loss feature is symmetric and can be well captured using a sum of gaussians
\begin{equation}
    N_\text{rel} (B) = N_a + \sum_i \frac{N_i}{\sigma_i \sqrt{2 \pi}} \exp{\left(\frac{-(B-B_\text{res}^{(i)})^2}{2\sigma_i^2}\right)} 
\end{equation}
to extract the resonances positions $B_\text{res}^{(i)}$, where $N_a$ is a constant offset accommodating for the background, $N_i$ scales the loss feature and $\sigma_i$ gives its width.
\\
If the modulation strength is sufficient and the precision of the magnetic offset field is good, Floquet-Feshbach resonances can be studied without additional modulation to broaden the resonances.
Then the asymmetric nature of the losses can be seen.
The lineshape of a Fano resonance 
\begin{equation}
    N_\text{rel} (B) = N_1 \frac{(q + \beta)^2}{1 + \beta^2} + N_a \: ,
\end{equation}
with
\begin{equation}
    \beta = \frac{B - B_\text{res}}{\Delta /2} \: ,
\end{equation}
captures the observed loss profile to a high degree.
The coefficient $N_1$ scales the resonance and $N_a$ captures a constant offset.
$B_\text{res}$ determines a resonance's position and $\Delta$ its width.
The parameter $q$ defines the shape of the resonance. 
For positive $q$ the loss minima is towards the side of higher magnetic field values of a resonance, while for negative $q$ it is towards lower values.
In the case $q=0$ the shape becomes symmetric.

\subsection{Analyzing the elastic response}
To analyze the response of the atomic cloud to the trap quench, we took several measurements for each data point, stacked the resulting absorption images and fitted the cloud's width with a Gaussian.
The time evolution of the so extracted widths was fitted using
\begin{equation}
    \sigma(t) = \sigma_0 + A e^{-\gamma_A t} \cos{(\omega t + \theta)} + B e^{-\gamma_B t} \:,
\end{equation}
where $\sigma_0$ is the initial width, $A$ and $\gamma_A$ the amplitude and decay rate of the oscillatory behavior with frequency $\omega$ and phase $\theta$. 
The last term captures the cloud's shrinkage due to non-oscillatory thermalization and trap losses with amplitude $B$ and decay rate $\gamma_B$.

\subsection{RF circuit}
\begin{figure}[b]
	\includegraphics{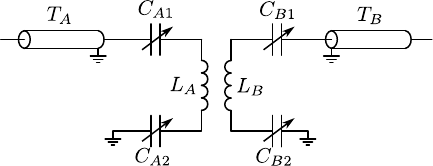}
	\caption[rf Circuit]{\textbf{Schematic of the rf circuit.} The circuit features two coupled, resonantly driven LC circuits, with variable capacitors for tuning the resonance frequency. Impedance matching to the $\qty{50}{\Omega}$ amplifiers is achieved using quarter-wavelength transformers, constructed by connecting several coaxial cables in parallel.}
	\label{fig:rf_circuit} 
\end{figure}

The strong rf magnetic field modulation is generated using a pair of rf Helmholtz coils placed inside a set of high-current coils that produce the bias magnetic field $B_0$. 
These rf coils are part of two resonantly driven LC circuits, as shown schematically in \autoref{fig:rf_circuit}.
To match the low impedance of the LC circuits at resonance with the $\qty{50}{\Omega}$ impedance of the amplifiers, we use quarter-wavelength transformers \cite{collin2000foundations} ($T_A$ and $T_B$ in \autoref{fig:rf_circuit}). 
These transformers are constructed by soldering several RG-58 $\qty{50}{\Omega}$ cables, cut to one quarter wavelength, in parallel.
Fine-tuning of the resonance frequency is achieved through variable capacitors in each LC circuit.

For single-frequency measurements, both LC circuits were tuned to the same frequency, whereas for the two-color drive, one circuit generated the first harmonic and the other the second harmonic.

\subsection{Calibrating the modulation strength}
For good agreement between theory and experimental data it is important to accurately measure the modulation strength produced by a given rf power at the place of the atoms.
Fortunately, the atoms themselves provide a precise way to measure the magnetic field modulation.
The rf dressing not only manipulates the scattering behavior of two atoms but also the energy levels of a single atom.
\\
The standard method to precisely measure the static magnetic field $B_0$ is to drive spin flip transitions between hyperfine levels and infer the magnetic field strength from the observed transition frequency.
The rf dressing leads to an AC Zeeman shift of the transition frequency and hence allows for the accurate measurement of the modulation strength, as demonstrated in \autoref{fig:ac_zeeman}.
\\
In addition, pickup coils are placed in the stray fields.
This provides a less accurate but still feasible way to measure the absolute modulation strength by comparing the induced voltage to numerical simulations to infer the magnetic field modulation strength at the atoms.
From Lenz's law the magnetic field modulation at the pickup coil is found to be
\begin{equation}
    B_\text{rf} = \frac{\sqrt{2}}{A 2 \pi \nu} \sqrt{\frac{P}{Z_0}} \:,
\end{equation}
where $P$ is the measured rf power, $Z_0$ the characteristic impedance (typically $Z_0=\qty{50}{\Omega}$).

\begin{figure}[b]
	\includegraphics{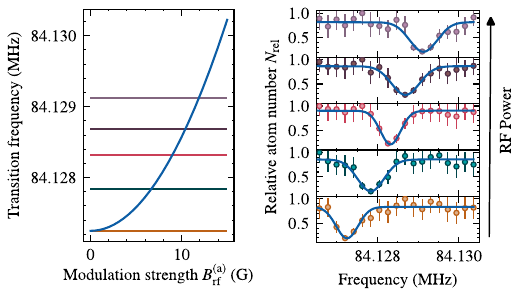}
	\caption[AC Zeeman shift single atom]{\textbf{AC Zeeman shift of the $\ket{b} \rightarrow \ket{c}$ transition.} The modulation shifts the transition frequency and hence allows for the accurate determination of the modulation strength with conventional methods of rf spectroscopy.}
	\label{fig:ac_zeeman} 
\end{figure}

\bibliography{bibliography}

\end{document}